\documentclass[aps,prb,showpacs,twocolumn,superscriptaddress]{revtex4}
\usepackage{amsmath}
\usepackage{amssymb}
\usepackage{epsfig}
\usepackage{color}
\usepackage{graphicx,amsmath}

\begin{document}
\title{Topological basis for understanding the
behavior of the heavy-fermion metal $\rm {\beta-YbAlB_4}$ under
application of magnetic field and pressure}
\author{V. R. Shaginyan}\email{vrshag@thd.pnpi.spb.ru}
\affiliation{Petersburg Nuclear Physics Institute, NRC Kurchatov
Institute, Gatchina, 188300, Russia}\affiliation{Clark Atlanta
University, Atlanta, GA 30314, USA}\author{A. Z. Msezane}
\affiliation{Clark Atlanta University, Atlanta, GA 30314, USA}
\author{K. G. Popov}\affiliation{Komi Science Center, Ural Division, RAS,
Syktyvkar, 167982, Russia}\affiliation{Department of Physics,
St.Petersburg State University, Russia}
\author{J.~W. Clark} \affiliation{McDonnell Center
for the Space Sciences \& Department of Physics, Washington
University, St.~Louis, MO 63130, USA} \affiliation{Centro de
Ci\^encias Matem\'aticas, Universidade de Madeira, 9000-390
Funchal, Madeira, Portugal}\author{V.~A. Khodel}
\affiliation{NRC Kurchatov Institute, Moscow, 123182, Russia}
\affiliation{McDonnell Center for the Space Sciences \&
Department of Physics, Washington University, St.~Louis, MO
63130, USA} \author{M.~V. Zverev} \affiliation{NRC Kurchatov
Institute, Moscow, 123182, Russia} \affiliation{Moscow Institute
of Physics and Technology, Dolgoprudny, Moscow District 141700,
Russia}
\begin{abstract}
Informative recent measurements on the heavy-fermion metal $\rm
\beta-YbAlB_4$ performed with applied magnetic field and
pressure as control parameters are analyzed with the goal of
establishing a sound theoretical explanation for the inferred
scaling laws and non-Fermi-liquid (NFL) behavior, which
demonstrate some unexpected features. Most notably, the
robustness of the NFL behavior of the thermodynamic properties
and of the anomalous $T^{3/2}$ temperature dependence of the
electrical resistivity under applied pressure $P$ in zero
magnetic field $B$ is at variance with the fragility of the NFL
phase under application of a field. We show that a consistent
topological basis for this combination of observations, as well
as the empirical scaling laws, may be found within
fermion-condensation theory in the emergence and destruction of
a flat band, {and explain that the paramagnetic NFL phase takes
place without magnetic criticality, thus not from quantum
critical fluctuations}. Schematic $T-B$ and $T-P$ phase diagrams
are presented to illuminate this scenario.
\end{abstract}
\pacs{71.27.+a, 71.10.Hf, 72.15.Eb}

\maketitle

\section{Introduction}

Recent measurements on the heavy-fermion (HF) metal $\rm
\beta-YbAlB_4$ have been performed under the application of both
a magnetic field $B$ and hydrostatic pressure $P$, with results
that have received considerable theoretical analysis
\cite{n2008,s2011,prb2011,conf2012,jpn2015,s2015,
col2009,col2012,quasi_jpn,jpn2014}.  Measurements of the
magnetization $M(B)$ at different temperatures $T$ reveal that
the magnetic susceptibility $\chi=M/B\propto T^{-1/2}$
demonstrates non-Fermi liquid (NFL) behavior and diverges as
$T\to0$, implying that the quasiparticle effective mass $m^*$
diverges as $m^*\propto B^{-1/2}\propto T^{-1/2}$ at a quantum
critical point (QCP).\cite{s2011} This kind of quantum
criticality is commonly attributed to scattering of electrons
off quantum critical fluctuations related to a magnetic
instability; yet in a single crystal of $\rm \beta-YbAlB_4$, the
QCP in question is located well away from a possible magnetic
instability, {making the NFL phase take place without magnetic
criticality. \cite{s2011}} Additionally, it is observed that the
QCP is robust under application of pressure $P$, in that the
divergent $T$ and $B$ dependencies of $\chi$ are conserved and
are accompanied by an anomalous $T^{3/2}$ dependence of the
electrical resistivity $\rho$.\cite{s2015} In contrast to
resilience of these divergences under pressure, application of
even a tiny magnetic field $B$ is sufficient to suppress them,
leading to Landau Fermi liquid (LFL) behavior at low
temperatures \cite{n2008,s2011}.  Thus, among other unusual
features, the metal $\rm \beta-YbAlB_4$ presents challenging
theoretical problems:  How to reconcile the frailty of its NFL
behavior under application of a magnetic field, with the
robustness of the NFL phase against application of pressure in
zero magnetic field; {How to explain that the paramagnetic NFL
phase takes place without magnetic criticality, thus not from
quantum critical fluctuations}.

\section{Scaling behavior}

To address these challenges within a topological scenario based
on the emergence of a fermion condensate (FC), we begin with an
examination of the scaling behavior of the thermodynamic
functions of this HF compound considered as homogeneous HF
liquid \cite{pr,book,quasi}. {We note that the existence of FC
has been convincingly demonstrated by purely theoretical and
experimental arguments, see e.g.
\cite{lids,lee,yudin,volovik,shash2014}}. The Landau functional
$E(n)$ representing the ground-state energy depends on the
quasiparticle momentum distribution $n_\sigma({\bf p})$. Near
the fermion-condensation quantum phase transition (FCQPT), the
effective mass $m^*$ is governed by the Landau equation
\cite{pr,book,land}
\begin{eqnarray}
\label{HC3} &&\frac{1}{m^*(T,B)}=\frac{1}{m^*(T=0,B=0)}\\&+&
\frac{1}{p_F^2}\sum_{\sigma_1}\int\frac{{\bf p}_F{\bf p_1}}{p_F}
F_{\sigma,\sigma_1}({\bf p_F},{\bf p}_1)\frac{\partial\delta
n_{\sigma_1}({T,B,\bf p}_1)} {\partial{\bf {p}}_1}\frac{d\bf
{p_1}}{(2\pi)^3},\nonumber
\end{eqnarray}
here written in terms of the deviation $\delta n_{\sigma}({\bf
p})\equiv n_{\sigma}({\bf p},T,B) -n_{\sigma}({\bf p},T=0,B=0)$
of the quasiparticle distribution from its field-free value
under zero pressure.  The Landau interaction $F({\bf p_1},{\bf
p_2})=\delta^2E/ \delta n({\bf p_1})\delta n({\bf p_2})$ serves
to bring the system to the FCQPT point where $m^*\to \infty$ at
$T=0$. As this occurs, the topology of the Fermi surface is
altered, with the effective mass $m^*$ acquiring temperature and
field dependencies such that the proportionalities $C/T \sim
\chi \sim m^*(T,B)$ relating the specific heat ratio $C/T$ and
the magnetic susceptibility $\chi$ to $m^*$ persist. Approaching
the FCQPT, $m^*(T=0,B=0) \to \infty$ and thus Eq.~\eqref{HC3}
becomes homogeneous, i.e., $m^*(T=0,B)\propto B^{-z}$ and
$m^*(T,B=0)\propto T^{-z}$, with $z$ depending on the analytical
properties of $F$.\cite{pr,book,obz94,quasi} On the ordered side
of the FCQPT at $T=0$, the single-particle spectrum
$\varepsilon({\bf p})$ becomes flat in some interval
$p_i<p_F<p_f$ surrounding the Fermi surface at $p_F$, coinciding
there with the chemical potential $\mu$,
\begin{equation}\label{flat}\varepsilon({\bf p})=\mu.
\end{equation}
At the FCQPT the flat interval shrinks, so that $p_i\to p_F\to
p_f$, and $\varepsilon({\bf p})$ acquires an inflection point at
$p_F$, with $\varepsilon({\bf p\simeq p_F}) -\mu\simeq
(p-p_F)^3$.  Another inflection point emerges in the case of a
non-analytical Landau interaction $F$, instead with
\begin{eqnarray}\label{var}
\varepsilon-\mu&\simeq&-(p_F-p)^2,p<p_F\\
\varepsilon-\mu&\simeq&\,\,\,(p-p_F)^2, p>p_F\nonumber
\end{eqnarray}
at which the effective mass diverges as $m^*(T\to 0)\propto
T^{-1/2}.$ Such specific features of $\varepsilon$ can be used
to identify the solutions of Eq.~\eqref{HC3} corresponding to
different experimental situations. In particular, the
experimental results obtained for $\rm \beta-YbAlB_4$ show that
near QCP at $B\simeq 0$, the magnetization obeys
\cite{n2008,s2011,prb2011,conf2012,jpn2015,s2015,col2009,col2012,quasi_jpn,jpn2014}
$M(B)\propto B^{-1/2}$.  This behavior corresponds to the
spectrum $\varepsilon({\bf p})$ given by Eq.~\eqref{var} with
$(p_f-p_i)/p_F\ll 1$.  At finite $B$ and $T$ near the FCQPT, the
solutions of Eq.~\eqref{HC3} determining the $T$ and $B$
dependencies of $m^*(T,B)$ can be well approximated by a simple
universal interpolating function \cite{pr,book,quasi}. The
interpolation occurs between the LFL ($m^* \propto a+bT^2$) and
NFL ($m^* \propto T^{-1/2}$) regimes separated by the crossover
region at which $m^*$ reaches its maximum value $m^*_N$ at
temperature $T_M$, and represents the universal scaling behavior
of
\begin{equation}\label{interp}
m_N^*(T_N)=\frac{m^*(T,B)}{m^*_M}=
\frac{1+c_2}{1+c_1}\frac{1+c_1T_N^2}{1+c_2T_N^{5/2}}.
\end{equation}
Here $c_1$ and $c_2$ are fitting parameters, $T_N=T/T_M$ is
the normalized temperature, and
\begin{equation}\label{MB}
m^*_M\propto B^{-1/2},
\end{equation} while
\begin{equation}\label{MT}
T_M\propto B^{1/2}\,\,\,{\rm and}\,\,\,T_M\propto B.
\end{equation}
It follows from Eqs.~\eqref{interp}, \eqref{MB}, and \eqref{MT}
that the effective mass exhibits the universal scaling behavior
\begin{equation}\label{interpm}m^*(T,B)= c_3\frac{1}{\sqrt{B}}m^*_N(T/B),
\end{equation}
with $c_3$ a constant.\cite{pr,book,quasi}
Eqs.~\eqref{interp}, \eqref{MB}, \eqref{MT}, and \eqref{interpm} will
be used along with Eq.~\eqref{HC3} to account for the experimental
observations on $\rm \beta-YbAlB_4$.  We note that the scaling behavior
at issue refers to temperatures $T\lesssim T_f$, where $T_f$ is the
temperature at which the influence of the QCP becomes negligible.
\cite{pr,book}.

\begin{figure} [! ht]
\begin{center}
\includegraphics [width=0.47\textwidth]{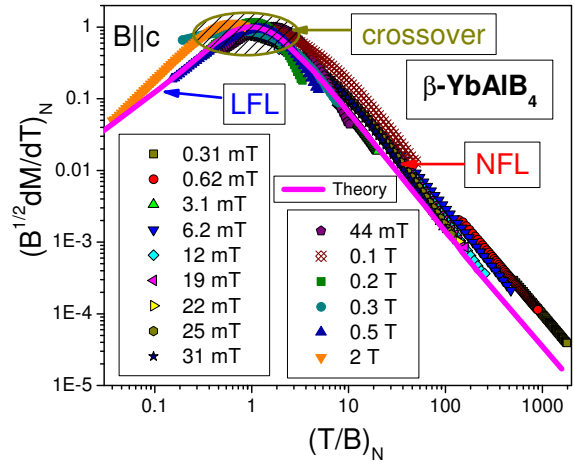}
\end{center}
\caption{(color online). Scaling behavior of dimensionless normalized
magnetization $(B^{1/2}dM(T,B)/dT)_N$ versus dimensionless normalized
$(T/B)_N$ at magnetic field values $B$ given in the legend.  Data
are extracted from the measurements described in Ref.~\onlinecite{s2015}.
Regions of LFL behavior, crossover, and NFL behavior are indicated
by arrows. The theoretical prediction is represented by a
single scaling function.}\label{fig1}
\end{figure}
Based on Eq.~\eqref{interpm}, we conclude the magnetization $M$
as described within the topological setting of fermion
condensation does exhibit the empirical scaling behavior, being
given by
\begin{equation}\label{TB}
M(T,B)=\int\chi(T,B_1)dB_1 \propto \int
\frac{m^*_N(T/B_1)}{\sqrt{B_1}}dB_1.
\end{equation}
At $T<B$ the system is predicted to show LFL behavior with
$M(B)\propto B^{-1/2}$, whereas at $T>B$, the system has entered the NFL
region and $M(T)\propto T^{-1/2}$. Moreover, $dM(T,B)/dT$
again exhibits the observed scaling behavior, with $dM(T,B)/dT\propto T$
at $T<B$ and $dM(T,B)/dT\propto T^{-3/2}$ at $T>B$.  Thus our analytical
results are in accordance experiment,\cite{s2011,conf2012,jpn2015}
free from fitting parameters and empirical functions.

\begin{figure} [! ht]
\begin{center}
\includegraphics [width=0.47\textwidth]{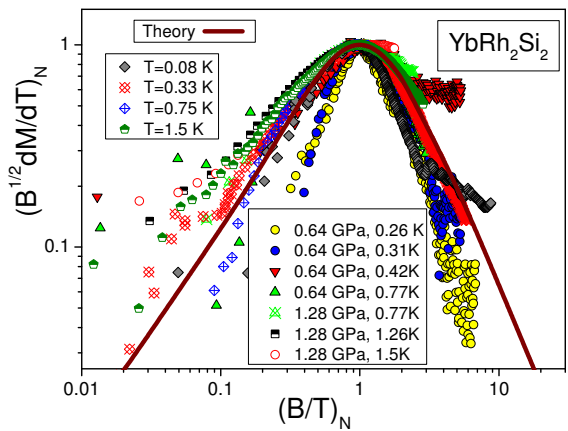}
\end{center}
\caption{(color online). Scaling behavior of the archetypal HF
metal $\rm YbRhSi_2$.  Data for $(B^{1/2}dM(T,B)/dT)_N$ versus
$(B/T)_N$ are extracted from measurements
of $dM/dT$ versus $B$ at fixed temperatures.\cite{mtprl,mtjpn}
The solid curve representing the theoretical calculation is
adapted from that of Fig.~\ref{fig1}. Applied pressures and
temperatures are shown in the legends.}
\label{fig2}
\end{figure}

In confirmation of the analysis of the scaling behavior,
Fig.~\ref{fig1} displays our calculations of the dimensionless normalized
magnetization measure $(B^{1/2}dM(T,B)/dT)_N$ versus the dimensionless
normalized ratio $(T/B)_N$. The normalization is implemented by dividing
$B^{1/2}dM(T,B)/dT$ and $T/B$ respectively by the maximum value of
$(B^{1/2}dM(T,B)/dT)_M$ and by the value of $(B/T)_M$ value the maximum
occurs. It is seen that the calculated single scaling function of the
ratio $(T/B)_N$ tracks the data over four decades of the normalized
quantity $(B^{1/2}dM(T,B)/dT)_N$, while the ratio itself varies over
five decades. It also follows from Eq.~\eqref{TB}
that $(B^{1/2}dM(T,B)/dT)_N$ exhibits the proper scaling behavior as a
function of $(B/T)_N$. Figure \ref{fig2} illustrates the scaling
behavior $(B^{1/2}dM(T,B)/dT)_N$ of the archetypal HF metal
$\rm YbRhSi_2$. The solid curve representing the theoretical
calculations is taken from Fig.~\ref{fig1}. Thus, we find that
the scaling behavior of $\rm \beta-YbAlB_4$, as extracted from
measurements \cite{mtprl,mtjpn} and shown in Fig.~\ref{fig1}, is
not unique, as Fig.~\ref{fig2} demonstrates the same crossover
under application of the magnetic field in the wide range of the
applied pressure.

\section{The Kadowaki-Woods ratio}

\begin{figure} [! ht]
\begin{center}
\includegraphics [width=0.47\textwidth]{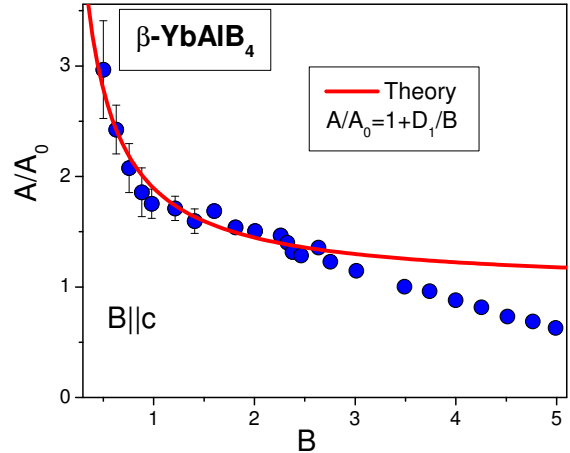}
\end{center}
\caption{(color online). Experimental data for normalized
coefficient $A(B)/A_0$ as represented by Eq.~(\ref{HFC}),
plotted as a function of magnetic field $B$ (solid circles).
Measured values of $A(B)$ are taken from
Ref.~\onlinecite{n2008}, with $D_1$ the only fitting parameter.
The solid curve is the theoretical prediction.} \label{fig3}
\end{figure}

Under application of magnetic fields $B>B_{c2}\simeq30$ mT and at
sufficiently low temperatures, $\rm \beta-YbAlB_4$ can be driven to
the LFL state having resistivity of the form $\rho(T)=\rho_0+AT^2$.
Measurements of the coefficient $A$ of the $T^2$ dependence have
provided information on its $B$-field dependence \cite{n2008}.
Being proportional to the quasiparticle-quasiparticle scattering
cross section, $A(B)$ is found to obey\cite{khodel:1997:B,gegenwart:2002}
$A\propto(m^*(B))^2$.  In accordance with Eq.~\eqref{MB}, this
implies that
\begin{equation}
A(B)\simeq A_0+\frac{D}{B},\label{ABDR}
\end{equation}
where $A_0$ and $D$ are fitting parameters.\cite{pr,book}  We
rewrite Eq.~\eqref{ABDR} in terms of the reduced variable $A/A_0$,
\begin{equation}
\frac{A(B)}{A_0}\simeq 1+\frac{D_1}{B},\label{HFC}
\end{equation}
where $D_1=D/A_0$ is a constant, thereby reducing $A(B)$ to a
function of the single variable $B$. Figure \ref{fig3} presents
the fit of $A(B)$ to the experimental data.\cite{n2008}
The theoretical dependence \eqref{HFC} agrees well with experiment
over a substantial range in $B$.  This concurrence suggests that the
physics underlying the field-induced re-entrance into LFL
behavior is the same for classes of HF metals.  It is important
to note here that deviations of the theoretical curve from the
experimental points at $B>2.5$ T are due to violation of the
scaling at the QCP.
\cite{jpn2015}
\begin{figure} [! ht]
\begin{center}
\includegraphics [width=0.47\textwidth]{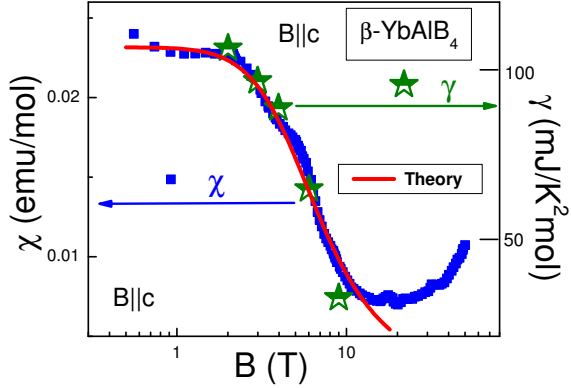}
\end{center}
\caption{(color online). Measurements\cite{jpn2015} of magnetic susceptibility
$\chi=dM/dB=a_1m^*_N$ (left axis, square data points) and electronic
specific heat coefficient $C/T=\gamma=a_2m^*_N$ (right axis,
stars), plotted versus magnetic field $B$.  Solid curve tracing
scaling behavior of $m^*_N$: theoretical results from present study
with fitting parameters $a_1$ and $a_2$.}\label{fig4}
\end{figure}

Fig.~\ref{fig4} compares our calculations of $\chi(B)\propto m^*$ and
$C/T=\gamma(B)\propto m^*$ with the experimental measurements.\cite{jpn2015}
Appealing to Eq.~\eqref{MB}, the behavior $A(B)\propto (m^*)^2$,
and the good agreement of theory with experiment shown in this
figure, we verify Eq.~\eqref{HFC} and conclude that the Kadowaki-Woods ratio
$A/\gamma^2\propto A/\chi^2 \simeq {\rm const.}$ is conserved in the case
of $\rm \beta-YbAlB_4$, much as in other heavy-fermion compounds
\cite{KW,pr,n2008,prb2011}.

\section{The phase diagrams}

\begin{figure}[!ht]
\begin{center}
\includegraphics [width=0.47\textwidth]{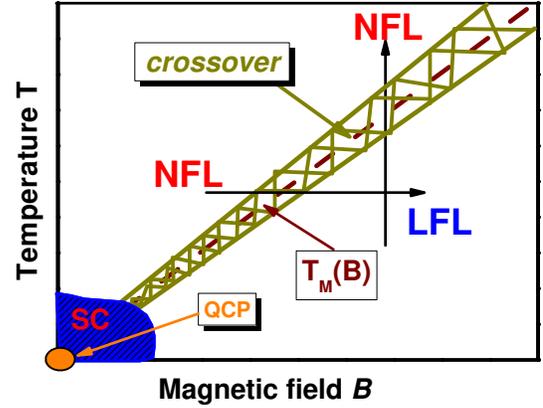}
\end{center}
\vspace*{-0.3cm} \caption{ Schematic $T-B$ phase diagram.
Vertical and horizontal arrows highlight LFL-NFL and NFL-LFL
transitions at fixed $B$ and $T$, respectively. Hatched area
separates the NFL phase from the weakly polarized LFL phase and
identifies the transition region. Dashed line in hatched
area represents the function $T_M\propto B$ (see Eq.~\eqref{MT}).
The QCP, located at the origin and indicated by the arrow, is
the quantum critical point at which the effective mass $m^*$
diverges. It is surrounded by the superconducting phase labeled
SC.} \label{fig5}
\end{figure}

The results of the above analysis of the scaling properties of
this HF system based on a topological scenario allow us to
construct the schematic $T-B$ phase diagram of
$\rm\beta-YbAlB_4$ presented in Fig.~\ref{fig5}, with the
magnetic field $B$ as control parameter.  At $B=0$, the system
acquires a flat band satisfying Eq.~\eqref{flat}, implying the
presence of a fermion condensate in a strongly degenerate state
of matter that becomes susceptible to transition into a
superconducting state \cite{khodel:2008,pr}. This NFL
fermion-condensate regime exists at elevated temperatures and
fixed magnetic field. QCP indicated by the arrow in
Fig.~\ref{fig5} is located at the origin of the phase diagram,
since application of any magnetic field destroys the flat band
and shifts the system into the LFL state, {\it provided} that
the superconducting state is not in play \cite{pr,book,sepl}.
The hatched area in the figure denotes the crossover region that
separates the NFL state from the LFL state, also indicated in
Fig.~\ref{fig1}.

\begin{figure}[!ht]
\begin{center}
\includegraphics [width=0.47\textwidth]{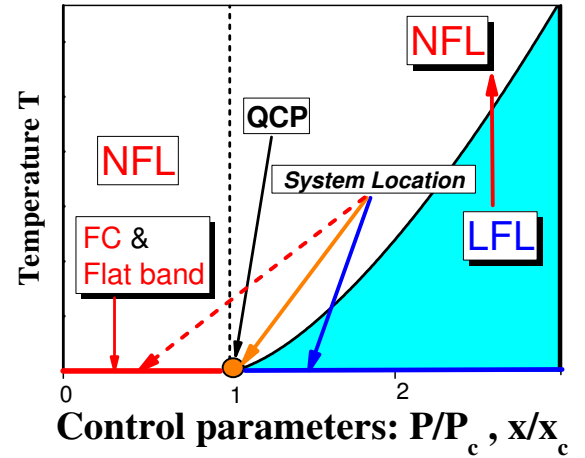}
\end{center}
\vspace*{-0.3cm} \caption{Schematic $T-x$ phase diagram of HF
system exhibiting a fermion condensate.  Pressure $P/P_c$ and
number-density index $x/x_c$ are taken as control parameters, with
$x_c$ the critical doping.  At $P/P_c>1$ and sufficiently low
temperatures, the system is located in the LFL state (shadowed area).
Moving past the QCP point to $P/Pc<1$ into the NFL region, the
system develops a flat band that is the signature of fermion
condensation (FC). The upward vertical arrow tracks the system moving
in the LFL-NFL direction along $T$ at fixed control parameters.
Not shown is the low-temperature stable phase satisfying the Nernst
theorem (superconducting in the case of $\rm \beta-YbAlB_4$)
that must exist for $P/P_c$ or $x/x_c$ below unity.} \label{fig6}
\end{figure}

Significantly, the heavy-fermion metal $\rm \beta-YbAlB_4$ is in
fact a superconductor on the ordered side of the corresponding
phase transition. When analyzing the NFL behavior of $\rho(T)$
on the disordered side of this transition, it should be kept in
mind that several bands simultaneously intersect the Fermi
surface, so that the HF band never covers the entire Fermi
surface.  Accordingly, it turns out that quasiparticles that do
not belong to the HF band make the main contribution to the
conductivity.  The resistivity therefore takes the form
$\rho(T)=m^*_{\rm norm}\gamma(T)$, where $m^*_{\rm norm}$ is the
average effective mass of normal quasiparticles and $\gamma(T)$
describes their damping. The main contribution to $\gamma(T)$
can be estimated as\cite{kss,kz1,cecoin5,JL2015} $\gamma\propto
T^2m^*(m^*_{norm})^2$. Based on Eqs.~\eqref{var} and \eqref{MT},
we obtain\cite{JL2015} $\rho(T)\propto T^{3/2}$.  On the other
hand, one would expect that at $T\to 0$ the flat band
\eqref{flat} comes into play, producing the behavior
$\rho(T)\propto A_1T$, with the factor $A_1$ proportional to the
flat-band range $(p_f-p_i)/p_F\ll1$.  However, such behavior is
not seen, because this area of the phase diagram is captured by
superconductivity, as already indicated in Fig.~\ref{fig5}.  The
low-$T$ resistivity $\rho(T,P=0)\propto T^{3/2}$ found
experimentally\cite{s2015} for the normal state of $\rm
\beta-YbAlB_4$ is consistent with this analysis. When the
pressure $P$ is raised to a critical value $P_c$, there is a
crossover to Landau-like behavior $\rho(T)=\rho_0+A_2T^2$.
Assuming that $P\propto x$, where $x$ is the doping or the HF
number density,\cite{mtjpn} we observe that such behavior
closely resembles the NFL behavior $\rho(T)\propto T^{1.5\pm
0.1}$ revealed in measurements of the resistivity in the
electron-doped high-$T_c$ superconductors $\rm
La_{2-x}Ce_xCuO_4$.\, \cite{armitage,greene1}  In that case the
effective mass $m^*(x)$ diverges as $x\to x_c$ or $P\to P_c$
\cite{armitage,greene1} according to\cite{pr,book,JL2015}
\begin{equation}\label{dx}
(m^*(x))^2\propto A\simeq
\left(a_1+\frac{a_2}{x/x_c-1}\right)^2.
\end{equation}
Here $a_1$ and $a_2$ are constants, while $x_c$ is the
critical doping at which the NFL behavior changes to LFL behavior,
the FC having decayed at $x_c$ and the system having moved to the
disordered side of the FCQPT.

In Fig.~\ref{fig6} we display the schematic $T-x$ phase diagram
exhibited by $\rm \beta-YbAlB_4$ when the system is tuned by
pressure $P$ or by number density $x$.  At $P/P_c<1$ (or
$x/x_c<1$) the system is located on the ordered side of
topological phase transition FCQPT and demonstrates NFL behavior
at $T\lesssim T_f$.  Thus, the NFL behavior induced by the FC
that persists at $P<P_c$ is robust under application of pressure
$P/P_c<1$.\cite{book,sepl} (We note that such behavior is also
observed in quasicrystals.\cite{quasi,QCM}). At low temperatures
the FC state possessing a flat band, highlighted in the figure,
is strongly degenerate.  This degeneracy stimulates the onset of
certain phase transitions and is thereby lifted before reaching
$T=0$, as required by the Nernst
theorem.\cite{khodel:2008,clark:2012} With rising pressure
(indicated by arrows in Fig.~\ref{fig6}), the system enters the
region $P/P_c>1$, where it is situated prior to the onset of the
FCQPT and demonstrates LFL behavior at sufficiently low
temperatures (shaded area in the figure). The temperature range
of this region shrinks when $P/P_c\to 1$, and $m^*$ diverges as
described by Eq.~\eqref{dx}. These observations are in accord
with the experimental evidence.\cite{s2015}

\section{Summary}

{To summarize, we have analyzed the thermodynamic properties of
the heavy-fermion metal $\rm \beta-YbAlB_4$ and explained their
enigmatic scaling behavior within a topological scenario in
which FC phase plays an essential role. We have explained why
the observed NFL behavior is extremely sensitive to a magnetic
field, and how the thermodynamic properties and anomalous
$T^{3/2}$ dependence of the electrical resistivity remain intact
under the application of a pressure}.

\section{Acknowledgments}

VRS acknowledges support from the Russian Science Foundation,
Grant No.~14-22-00281. This research was also supported by RFBR
Grants\#14-02-00044 and 15-02-06261, and by grant NS-932.2014.2
from the Russian Ministry of Sciences (MVZ and VAK).  VAK and
JWC thank the McDonnell Center for the Space Sciences for timely
support.

\end{document}